\documentclass[showpacs,twocolumn,aps,preprintnumbers,letterpaper]{revtex4}
\usepackage{amsmath,amssymb}
\usepackage{epsfig}
\usepackage{graphicx}
\usepackage{amsmath}
\usepackage{slashed}
\usepackage{amsfonts}
\usepackage{epstopdf}
\usepackage{color}

\newcommand{\be}{\begin{equation}}
\newcommand{\ee}{\end{equation}}
\newcommand{\bear}{\begin{eqnarray}}
\newcommand{\eear}{\end{eqnarray}}
\newcommand{\ba}{\begin{array}}
\newcommand{\ea}{\end{array}}

\def\be{\begin{eqnarray}}
\def\ee{\end{eqnarray}}
\def\bea{\be}
\def\eea{\ee}

\def\roughly#1{\mathrel{\raise.3ex\hbox{$#1$\kern-.75em%
\lower1ex\hbox{$\sim$}}}}

\begin{document}

\title{Pion Condensation by Rotation in a Magnetic field}

\author{Yizhuang Liu  and Ismail Zahed}
\email{yizhuang.liu@stonybrook.edu}
\email{ismail.zahed@stonybrook.edu}
\affiliation{Department of Physics and Astronomy, Stony Brook University, Stony Brook, New York 11794-3800, USA}



\date{\today}
\begin{abstract}
We show that the combined effects of a rotation plus a magnetic field can cause  charged pion condensation. 
 We suggest that this phenomenon may yield to observable effects in
current  heavy ion collisions at collider energies, where large magnetism and rotations are expected in off-central
collisions.
\end{abstract}


\maketitle

\setcounter{footnote}{0}


\section{Introduction}

The combined effects of rotations and magnetic fields on  Dirac fermions are realized in a
wide range of physical settings ranging from macroscopic spinning neutron stars and black holes
\cite{VILENKIN}, all the way to
microscopic anomalous transport in Weyl metals~\cite{WEYL}. In any dimensions, strong magnetic fields
reorganize the fermionic spectra into Landau levels, each with a huge planar degeneracy that
is lifted when a paralell rotation is applied. The past decade has seen a large interest in the
chiral and vortical effects and their relationship with anomalies~\cite{ANOMALY} (and references therein).

Perhaps, a less well known effect stems from the dual combination of a rotation and magnetic
field on free or interacting Dirac fermions. Recently, it was noted  that this dual combination
could lead to novel effects for composite fermions at half filling in 1+2 dimensions under the
assumption that {they are Dirac fermions}~\cite{US},  and more explicitly for free and interacting
Dirac fermions in 1+3  dimensions~\cite{YIN,LIAO,FUKU}.  Indeed, when a rotation is applied along a magnetic field,
the charge density was observed to increase {\it in the absence} of a chemical potential.
A possible relationship of this phenomenon to
the Chern-Simons term in odd dimensions, and the chiral anomaly in even dimensions was
suggested.
Recently, there have been few studies along these lines using effective models of the NJL type
in 1+3 dimensions, where the phenomenon of charge density enhancement was also confirmed
with new observations~\cite{LIAO,FUKU,MAXIM}.

Current heavy ion collisions at collider energies  in non-central collisions involve large angular momenta
in the range $10^3-10^5\,{\hbar }$~\cite{BECATINI,FENG}. Recently, STAR reported a large vorticity with
$\Omega\sim (9\pm 1)\,10^{21}\,s^{-1}\sim 0.05\,m_\pi$, by measuring the global polarization  of $\Lambda$ and
$\bar\Lambda$ in off central AuAu collisions in the Beam Energy Scan program~\cite{STAR}. During the prompt
part of the collision, large magnetic fields $B\sim m_\pi^2$ are expected~\cite{DIMA}. In this letter, we show that the
combined effects of magnetism plus a rotation can induce a pion superfluid phase in off-central heavy ion collision.
This superfluid phase maybe at the origin of the large multi-pion correlations reported by ALICE~\cite{ALICE}, as
also suggested by a recent non-equilibrium study~\cite{BEGUN}.

In section II we show how this combination yields a charged pion condensation.
In section III, we make an estimate of the amount of pion condensation in current
heavy ion collisions at collider energies. Our conclusions are in section V.

\begin{figure}[h]
  \begin{center}
\includegraphics[width=7cm]{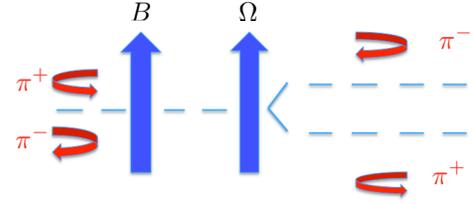}
  \caption{Under the action of an external magnetic field $B$ the $\pi^\pm$ undergo opposite rotations in the  Lowest Landau Level (LLL) which is degenerate. The action of a parallel rotation ($\vec\Omega\cdot\vec B>0$) lifts the degeneracy in the LLL.  The energy of the $\pi^+$ shifts down and splits away from the energy of the $\pi^-$ that shifts up. The action of an anti-parallel rotation  
  ($\vec\Omega\cdot\vec B<0$)  exchanges the role of $\pi^+$  and $\pi^-$.}
 \label{fig_bomega}
 \end{center}
\end{figure}

\section{Pion condensation}

In the presence of a fixed magnetic field in the +z direction ${\bf B}=B\hat z$, the charged
$\pi^\pm$ pion spectrum is characterized by highly degenerate Landau Levels (LL)

\be
\label{SZ1}
{E}_{np}=\left(|eB|(2n+1)+p^2+m_\pi^2\right)^{\frac 12}
\ee
with $p$ the pion momentum along the 3-direction, each with a degeneracy
$N=|eB|S/2\pi$ with $S=\pi R^2$ the area of the plane transverse to $B$. 
We will assume that the magnetic length $l_M=1/\sqrt{|eB|}\ll R$ for the LL
to fit within $S$. In 
the circular gauge the degeneracies of the LL are identified with the eigenstates of the
z-component of the angular momentum in position space. They are labeled by $l$ 
which enters the azimuthal wave-function as $e^{il\varphi}$ with the restriction
$-n\leq l\leq N-n$ where $n$ labels the LL. For the Lowest Landau Level (LLL) 
with $n=0$, $l$ has a fixed sign since $0\leq l\leq N$. After quantization, the
angular momentum for positive charged particles is $l$ and for negative charged
particles is $-l$. This means that in the LLL, the  $\pi^+$ spins along the magnetic field, 
while the $\pi^-$ spins opposite to the magnetic field as illustrated in Fig.~\ref{fig_bomega}.  

When a rotation $\Omega$ along the magnetic field is applied,  it causes the spectrum to shift linearly.  
Throughout we will consider the parallel case with $\vec\Omega\cdot\vec B>0$ unless specified
otherwise. With this in mind, and in the rotating frame

\be
\label{SZ2}
{E}_{np}\rightarrow {E}_{np}-{\Omega}{L_z} \equiv {E}_{np}-j{\Omega}{l}
\ee
with $j=+1$ for positively charged pions (particles) and  $j=-1$ for negatively charged pions
(anti-particles). As a result, the degeneracy of each LL is lifted.
In particular, the  $\pi^+$  in the LLL splits down and the  $\pi^-$ in the 
LLL splits up as also illustrated in Fig.~\ref{fig_bomega}. Since the
chargeless pions $\pi^0$ are unaffected by the magnetic field, their rotational shift averages out.
Also we note that causality requires $v=\Omega R\leq 1$~\cite{FUKU} which together with the magnetic
length constraint (see above) translates to $l_M\ll R<1/\Omega$.

The mechanism of $\pi^\pm$ splitting by a rotation parallel to a magnetic field  in the LLL can cause $\pi^+$ pion
condensation. Indeed, in the shifted spectrum (\ref{SZ2}), the combination $\mu_l=\Omega l$ plays the role of
a chemical potential for $\pi^+$ and $-\mu_l=-\Omega l$ for $\pi^-$, in much the same way as noted for fermionic
particles and anti-particles in the LLL~\cite{VILENKIN,FUKU,VOLO,US2}. Therefore, when $\mu_N=N\Omega$ 
apparently exceeds the $\pi^+$ effective mass in the LLL, $m_0=\sqrt{eB+m_\pi^2}$, but is still below  the $\pi^+$ effective mass in  the first LL  with $n=1$, the LLL  $\pi^+$  may Bose condense, provided that charge conservation is enforced. 

For a fixed and isolated volume $V=SL$  with no charge allowed to flow in or out, strict charge conservation 
in the co-moving frame is achieved by introducing a charge chemical potential $\mu$,  in addition to the 
induced chemical potential  $j\Omega l$ by rotation. 
(For an open volume discussion see~\cite{US2} and references therein). For the LLL,  charge conservation requires that the number of $\pi^\pm$
in $V$ are equal at any temperature 

\bea
\label{P1}
&&\sum_{l=0}^{N}\int\frac{dp}{2\pi}\frac{1}{e^{\frac 1T (E_{0p}-l\Omega-\mu)}-1}=\nonumber\\
&&\sum_{l=0}^{N}\int\frac{dp}{2\pi}\frac{1}{e^{\frac 1T (E_{0p}+l\Omega+\mu)}-1}
\eea
This equation is solved by inspection with $\mu=-\frac{N\Omega}{2}$. Therefore, the orbital assignments 
$l=N-m$ and $l=m$  for $\pi^{+}$ and $\pi^{-}$ in the LLL will have the same ocupation number 

\bea
\label{P2}
&&n_{\pi^+}(l=N-m)=n_{\pi^-}(l=m)\nonumber \\&&= \int\frac{dp}{2\pi}\frac{1}{e^{\frac 1T (E_{0p}-N\Omega/2+l\Omega)}-1}
\eea
with $0\le m \le N$.
For $N\Omega>2m_0$  simultaneous condensation occurs for $m=0$, i.e. $\pi^+$ with  $l=N$ and $\pi^-$ with $l=0$. 
For $(N-2)\Omega>2m_0$ the condensation involves both $m=0,1$. As we increase $\Omega$ such that
$\Omega=2m_0$, all $m\leq \frac N2$
will condense, i.e. $\pi^+$ with $\frac{N}{2}\le l\le N$ and $\pi^-$ with $0 \le l\le\frac{N}{2}$, and so on.


Now consider the rotating ground state with $T=0$ and $N\Omega>2m_0$ but $(N-2)\Omega<2m_0$, so that
only the $l=N$ state for $\pi^{+}$ and $l=0$ state for $\pi^{-}$ condense. The energy per  unit length  in the 
Bose-Einstein condensate (BEC) state is

\be
\label{P3}
{\cal E}_{\pi\Omega}=-{\bf n}\,(N\Omega-2m_0)+d_N{\bf n}^2
\ee
with the Coulomb factor

\be
d_N\approx \frac {e^2}{2}  \int_{l_M}^{R} 2\pi rdr \left( \frac{1}{2\pi r}\right)^2=\frac{e^2}{4\pi}\ln \frac{R}{a}\approx \frac{e^2}{8\pi} \ln N
\ee
$d_N$ characterizes 
 the electric field energy stored between two charged rings with radius $l_M\sim 1/\sqrt{eB}$ and charge $-e$ ($\pi^-$),
and radius $R\gg l_M$ and charge $+e$ ($\pi^+$). The Coulomb self-energy is subleading and omitted.
In the ground state, the BEC density ${\bf n}$ is fixed by  minimizing the energy density ${\cal E}_{\pi \Omega}$ in (\ref{P3}),
with the result

\bea
\label{PCHARGE}
{\bf n}=\theta(N\Omega-2m_0)\frac{N\Omega-2m_0}{2d_N} 
\eea





The rotating $\pi^+$ condensate induces a uniform
magnetic field ${\bf b}_z$ that enhances the applied initial field $B$, and back-reacts on the 
formation of the charged condensates to order $\alpha=e^2/4\pi$. Indeed, the  rotating  BEC
of $\pi^+$  at $r=R$  generates an azimutal current

\be
J^{\theta}[{\bf n}]=\frac{e{\bf{n}}N}{m_0r}|f_{0N}|^2\approx\frac{e^2B{\bf n}}{4\pi m_0}\delta(r-R)
\ee
where $f_{0N}$ is the LLL with angular momentum $l=N$. The corresponding induced magnetic field 

\be
{\bf b}_z[{\bf n}]=\frac{e^2B{\bf n}}{4\pi m_0}
\ee
modifies the applied magnetic field to order $\alpha=e^2/4\pi$ through $B\rightarrow B+{\bf b}_z[n]$.
The back-reacted LL problem amounts to the following substitutions for $m_0$ and $N$

\bea
m_0^2[{\bf{n}}]&=&m_\pi^2+eB\left(1+\frac{e^2{\bf n}}{4\pi m_0}\right)\nonumber\\
N[{\bf n}]&=&N\left(1+\frac{e^2{\bf n}}{4\pi m_0}\right)
\eea
The back-reacted density for the $\pi^\pm$ condensates follows by minimizing the energy per unit length

\bea
\label{EBACK}
&&{\cal E}[\Omega,{\bf n}]=\nonumber \\ 
&&-{\bf n}(N[{\bf  n}]\Omega-2m_0[{\bf  n}])+{\bf  n}^2e^2\left(\frac{eBN}{16 \pi m_0^2[{\bf n}]}+\frac{\ln N[{\bf n}]}{8\pi}\right)
\nonumber\\
\eea
This is the analogue of (\ref{P3}) with $d_N=\frac{e^2\ln N({\bf n})}{8\pi}$,  including the additional magnetic energy from the back reaction 

\be
\pi R^2\,\frac{{\bf b}_z^2}{2}=\frac{{\bf{n}}^2e^4B^2R^2}{32\pi m_0^2[{\bf n}]}=\frac{e^3BN{\bf{n}}^2}{16 \pi m_0^2[{\bf n}]}
\ee
The true ground state follows by minimizing (\ref{EBACK}) with respect to ${\bf n}$.
Both $m_0[\bf{n}]$ and  $N[\bf{n}]$ are observed to be weakly dependent on the ${\bf n}$-contributions from the back-reaction. 
We now explore the physical implication of (\ref{EBACK}) in heavy ion collisions.





\section{Pion BEC  in heavy-ion collisions}

\begin{figure}[h]
  \begin{center}
\includegraphics[width=9cm]{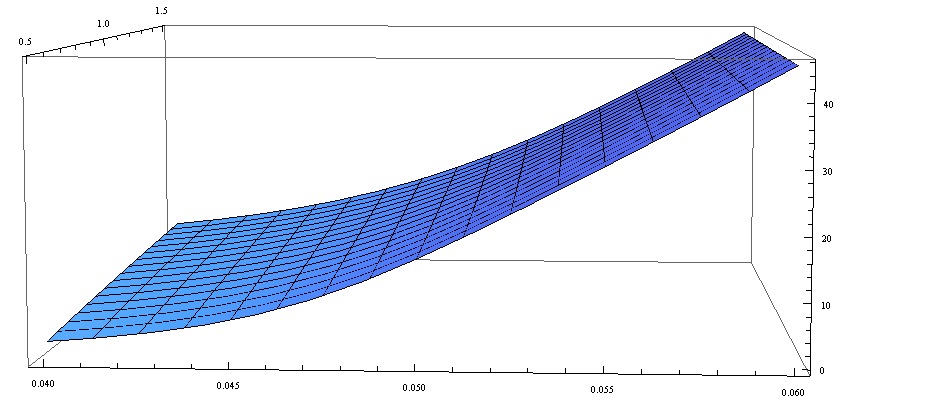}
  \caption{The mean number of condensed  pions ${\mathbb N}_{\pi^+}={\mathbb N}_{\pi^-}$ in the range $0.04\,m_{\pi}\le \Omega\le 0.06\,m_{\pi}$, for $\mu_f=0.86\,m_{\pi}$ and $0.5\,m_\pi\le T\le 1.5\,m_\pi$.}
    \label{fig_massstrong}
 \end{center}
\end{figure}

Current heavy ion collisions at collider energies are characterized by large
angular momenta  $l\sim 10^3-10^5\,{\hbar }$~\cite{STAR} and large magnetic fields
$B\sim m_\pi^2$~\cite{DIMA} in off central collisions. Assuming that
at chemical freeze-out, $R\sim 10$ fm with still $eB\sim m_\pi^2$, this would translates to a LL degeneracy
$N=eBR^2/2\sim (m_\pi\times10\,{\rm fm})^2\sim 100/4$ and a rotational chemical potential $\mu_N=N\Omega\sim 1.25\,m_\pi$.
From the hadro-chemistry  analysis, the  pion chemical
 potentials at freeze-out are typically $\mu_f\sim 0.5\,m_\pi$ at RHIC,  and $\mu_f\sim 0.86\,m_\pi$ at the LHC~\cite{BEGUN}.
 With the rotation at finite $B$, they would translate to $\mu_\pi=\mu_N+2\mu_f\sim 1.96\,m_\pi$  and
 $2.98\,m_\pi$  respectively. Since the threshold of the LLL for the combined $\pi^\pm$ pion energy is $2\sqrt{2}\,m_\pi$,
charge pion condensation is possible. Using (\ref{EBACK}) at finite $T,\mu_f$, the number of $\pi^{\pm}$ pions in the BEC are

\be
\mathbb N_{\pi^\pm}=\frac{\sum_{n=0}^{\infty}n \,e^{-\frac 1T(L{\cal E}[\Omega,\frac{n}{L}]-2n\mu_f)}}
{\sum_{n=0}^{\infty}e^{-\frac 1T(L{\cal E}[\Omega,\frac{n}{L}]-2n\mu_f)}}
\ee  
For  $L\sim 10$ fm, $eB\sim m_\pi^2$ and  $N\approx 25$, we show in Fig.~\ref{fig_massstrong} the average number of
condensed $\pi^\pm$  for temperatures in the range $0.5\,m_\pi\le T\le 1.5\,m_\pi$ and rotations in the range
$0.04\,m_{\pi}\le \Omega\le 0.06\,m_{\pi}$ for the most favorable case with $\mu_f=0.86\, m_\pi$
at the LHC.  It is interesting to note that the ALICE collaboration has recently reported a large coherent emission from multi-pion correlation studies  in Pb-Pb collisions~\cite{ABELEV}.


\section{Conclusions}

The combined effects of a rotation parallell to a magnetic field yields to pion condensation both
in the vacuum and at finite temperature. The $\pi^+$ condense at the edge, wile the $\pi^-$ at
the center in equal amount when charge conservation is strictly enforced in a closed volume.
Since parallell rotations and magnetic fields can be generated in current heavy ion collisions
at collider energies, charged pion condensation could be generated if the combined effects survive
with considerable strength in the freeze-out phase. Such effects are likely 
to affect both the flow of charge particles and their number fluctuations.  This separation of charged
bosons by centrifugation in a magnetic field may also be probed in atomic physics (trapped and cooled atoms), in 
condensed matter physics (quantum Hall effect) and possibly compact stars (magnestars).

\section{\label{acknowledgements} Acknowledgements}

We thank Edward Shuryak for a discussion.
 This work was supported in part by the U.S. Department of Energy under Contract No.
DE-FG-88ER40388.

\section{Appendix:  Klein-Gordon equation in rotating frame}

We now present and explicit derivation of the Klein-Gordon spectrum in a rotating frame.
In particular, we will recover the assignment given in (\ref{SZ2}) in the infinite volume limit. The
finite volume effects will be briefly discussed. We will also show that the appearance of the electric field 
in the rotating frame cancels out in the pion spectrum. No net force is generated by a frame change.

Consider the metric for a rotating frame in 1+3-dimensions with mostly negative signature $(+,-,-,-)$

\bea
\label{A1}
ds^2=(1-\Omega^2\rho^2)dt^2+2y\Omega dxdt-2x\Omega dydt-dr^2
\eea
The co-moving frame is defined as $e_a=e^{\mu}_{a}\partial_{\mu}$ with
$(e_0,e_1,e_2,e_3)=(\partial_t+y\Omega\partial_x-x\Omega\partial_y,\partial_x,\partial_y,\partial_z)$.
Now,  consider a constant magnetic field in the z-direction $B\hat z$ as described by
the circular vector potential $A_{R\mu}=B(0,y_R/2,-x_R/2,0)$ in the rest frame 
sub-labeled by $R$.
The coordinate transformation to the rotating frame $r_{R}=r,t_R=t,\theta_{R}=\theta+\Omega t$
allows the re-writing of the vector potential  in the {\it rotating} frame as

\bea
A_{\mu}=B(-\Omega r^2/2,y/2,-x/2,0)
\label{A4}
\eea
In the rotating frame there is in addition to the magnetic field $B\hat z$, an induced electric field 
$\vec E=\Omega B\vec r$. This is expected from a Lorentz transformation from the fixed frame with
$B\hat z$ to the co-moving frame with $B\hat z$ and $\vec E$. 

A charged scalar field $\Pi$ in the rotating frame and subject to the vector potential (\ref{A4}) is characterized by the
Lagrangian

\be
\label{A5}
{\cal L}=|(D_t+y\Omega D_x-x\Omega D_y)\Pi|^2-|D_i\Pi|^2-m_\pi^2\Pi^{\dagger}\Pi
\ee
with the long derivative $D_\mu=\partial_{\mu}+ieA_{\mu}$. We now note the identity

\be
\label{A55}
D_t+y\Omega D_x-x\Omega D_y=\partial_t+y\Omega \partial_x-x\Omega \partial_y
\ee
The electric field following from (\ref{A4}) cancels out. The co-moving
frame corresponds only to a frame change with no new force expected. 
With this in mind, the equation of motion for the charged field in the rotating frame is

\be
\label{A6}
-(\partial_t+y\Omega \partial_x-x\Omega \partial_y)^2\Pi-D_i^{\dagger}D_i\Pi-m_\pi^2\Pi=0
\ee
In the infinite volume case, we solve (\ref{A6}) using the algebraic ladder construction with

\bea
\label{A7}
a=&&\frac{i}{\sqrt{2eB}}(D_x+iD_y)\nonumber\\
b=&&\frac{1}{\sqrt{2eB}}(2\partial +\frac{eB}{2}\bar z)
\eea
Choosing the positive z-direction to be that for which $eB$ is positive, yields the operator
identities

\bea
\label{A8}
&&D_x^{\dagger}D_x+D_y^{\dagger}D_y=eB(2a^{\dagger}a+1)\nonumber\\
&&L_z=i(-x\partial_y+y\partial_x)=b^{\dagger}b-a^{\dagger}a
\eea
The general stationary solution to (\ref{A6}) is of the form $\Pi=e^{ipz-iEt}f$ with $f$ solving

\be
\label{A9}
(E+\Omega L_{z})^2 f=(m_\pi^2+p^2)f+eB(2a^{\dagger}a+1)f
\ee
The normalizable solutions form a tower of LL 

\bea
\label{A10}
&&f_{mn}=\frac{1}{\sqrt{m!n!}}(a^{\dagger})^n(b^{\dagger})^{m}f_{00}\nonumber\\
&&(E_{mn}+\Omega(m-n))^2=eB(2n+1)+m_\pi^2\equiv E_n^2
\eea
with $f_{00}\sim  e^{-\frac{eB}4(x^2+y^2)}$ as the LLL. For the LL to fit in a volume $V=LS$, we need
$m, n\leq N$.  The quantized charged field $\Pi$ in the rotating frame is

\be
\label{A11}
\Pi= \int \frac{dp}{2\pi }\sum_{nm}\frac{f_{mn}}{\sqrt{2E_n}}(a_{nmp}e^{-iE^{+}t+ipz}+b^{\dagger}_{nmp}e^{iE^{-}t-ipz})\nonumber\\
\eea
with the bosonic canonical rules

\be
\label{A12}
  \left[b_{nmp},b^{\dagger}_{n^{\prime}m^{\prime}p^{\prime}}\right]=\left[a_{nmp},a^{\dagger}_{n^{\prime}m^{\prime}p^{\prime}}\right]=2\pi\delta_{nn^{\prime}}\delta_{mm^{\prime}}\delta(p-p^{\prime})\nonumber
\ee
The particle state created by $a^{\dagger}_{nmp}$ has energy $E^{+}=E_n-\Omega(m-n)$ , charge $+e$ and
orbital  angular momentum in the z direction as $m-n$. The  anti-particle state created by $b^{\dagger}_{nmp}$ has 
 energy $E^{-}=E_n+\Omega(m-n)$ , charge $-e$ and orbital angular momentum in the z direction as $-m+n$.  
 Therefore, the energy relationship between the rotating frame and the rest frame is
$E=E_R-\Omega L_z$. This is in agreement with (\ref{SZ2}) where we have set $l=m-n$ and
defined $L_z=jl$ with $j=+1$ for particle or positive charge state and $j=-1$ for anti-particle or negative charge state.

In a finite volume (\ref{A6}) can be solved using instead the circular wavefunctions with zero boundary conditions

\be
\label{A13}
f_l(r,\theta)=e^{il\theta}r^{|l|}e^{-\frac{eBr^2}{4}}{}_1F_1\left(-a,|l|+1,\frac{eBr^2}{2}\right)
\ee
where ${}_1F_1$ is a hypergeometrical function with the parameter

\be
\label{A14}
-a(l)=\frac{1}{2}(|l|-l+1)-\frac{1}{2eB}((E+\Omega l)^2-p^2-m_\pi^2)
\ee
Thus, for positve angular momentum states we have

\bea
\label{A15}
&&(E+\Omega l)^2=p^2+m_\pi^2+eB(2a(l)+1)\nonumber\\
&&{}_1F_1(-a (l),l+1,\frac{eBR^2}{2})=0
\eea
The zero of the hypergeometric function fixes $a(l)$, and therefore the LL  for a finite volume. 
 For example, for $N=25$  we have $l=20$,  and  $a_{min}(20)=0.43$ for which the energy is $E=\sqrt{m_\pi^2+1.86eB}-20\Omega$. For $eB=m_\pi^2$, the treshold rotation is $\Omega_c=\frac{1.69}{20}\sqrt{eB}$.  Note that for $N=25$, we have $R=\sqrt{50/eB}$ and the luminal constraint is still fulfilled since $\Omega_c R=0.59<1$. For $N=100$ we have $l=84$,  and  $a_{min}(84)=0.18$  for which the energy is  $\sqrt{m_\pi^2+1.36eB}-84\Omega$ with the threshold $\Omega_c R=0.25$. For $N=1000$ 
 we have  $l=935$ and  $a_{min}(935)<0.1eB$. Thus as $N$ goes to infinity, the state with the lowest energy will aproach $l=N$, and our aproximation  in the main text becomes more precise.  Note that for exactly $l=N-1$, we always have 
 $a_{min}(N-1)=1$, and the energy for such state is  $\sqrt{m_\pi^2+3eB}-(N-1)\Omega$



 \vfil

\end{document}